\documentclass[aps,prl,a4paper,10pt,notitlepage,footinbib,superscriptaddress,showpacs,floatfix]{revtex4-1}%
\usepackage[english]{babel}
\usepackage[T1]{fontenc}
\usepackage{amssymb,amsmath,amsfonts}
\usepackage{textcomp}
\usepackage{graphicx,color}
\usepackage[dvipsnames]{xcolor}
\usepackage[utf8x]{inputenc}
\usepackage{bm}
\usepackage{float}
\usepackage{comment}

\usepackage{tabularx}

\usepackage{xr}
\begin{document} 

\title{Design Principles for Programmable Topological Soft Networks}

\author{Andrea Bonato}
\author{Davide Marenduzzo}
\affiliation{SUPA, School of Physics and Astronomy, The University of Edinburgh, Edinburgh, EH9 3FD, Scotland, United Kingdom}
\author{Enzo Orlandini}
\affiliation{Department of Physics and Astronomy, University of Padova and Sezione INFN, Padova, Via Marzolo 8, I-35131 Padova, Italy
}

\begin{abstract}
Physical crosslinks in conventional gels are static and spatially localised, fundamentally limiting their capacity to adapt, self-heal, or perform mechanical work. Topological gels, which are held together by physical entanglements such as threadings, links, and mechanical interlocks, constitute a different paradigm by treating connectivity as a mobile, dynamic degree of freedom. Here, by using molecular dynamics simulations of semiflexible polymers and threadable colloidal rings, we establish some general design principles that transform self-assembled topological gels into programmable soft materials. We present three different modes of control.
First, by controlling the mobility of the entanglements, one can tune the mechanical response of the material. Quenched anchoring forces the network to store elastic stress and yield, whereas with annealed anchoring the material can relax stresses by redistributing its entanglements whilst preserving global connectivity. 
Second, by controlling the degree or type of geometric confinement, it is possible to template macroscopic architecture, converting for instance 3D isotropic gels into surface-spanning 2D nets or quasi-1D topological rings with distinct graph-spectral signatures.
Third, by allowing dynamical rewiring of entanglements through polymer reconnection, the material topology can be rewritten after assembly, yielding interlocked structures dominated by simple pairwise links rather than the highly complex linked states previously observed in reconnecting flexible polymers. 
By establishing a direct link between microscopic entanglement dynamics and emerging mechanical function, our work offers design principles for adaptive soft metamaterials and provides insights into biological topological networks such as kinetoplast DNA and 
chromatin loop networks.
\end{abstract}
\maketitle

\section{Introduction}

The properties of standard polymer networks are governed by the density, strength, and spatial distribution of their chemical crosslinks. Topological materials, in contrast, derive their integrity from non-covalent topological constraints, which we shall hereafter refer to as {\it topological entanglements}, which are neither localised at specific chemical bonds nor permanently fixed in space. Indeed physical threadings, links, and mechanically interlocked junctions can migrate through the network upon deformation and, in some cases, can be created or destroyed dynamically.

Topological interactions are fundamental to understanding the mechanical properties of polymer melts~\cite{rosa2014ring,sakaue2011ring} and are known to generate unusual mechanical behaviour, such as non-trivial elasticity in Olympic gels~\cite{raphael1997olympic,lang2014swelling,krajina2018} or extreme extensibility via the ``pulley effect'' in slide-ring polyrotaxanes~\cite{noda2014topological,hart2021nrm}.
However, the network topology in existing synthetic systems is largely hardwired during chemical synthesis~\cite{oconnor2020thickening}. Conversely, topological self-assembly allows long-lived entanglements to form spontaneously under concentration or confinement~\cite{micheletti2024topology}. Biological systems exploit this extensively: kinetoplast DNA (kDNA) forms a natural 2D chainmail network of interlocked minicircles~\cite{he2023kinetoplast,Klotz2020,michieletto2015kinetoplast}, while protein rings like cohesin act as mobile slip-links that thread and organise nuclear chromatin loops. Computer simulations have shown that self-assembly of reconnecting polymers produces equilibrium gels rich in complex knots and links~\cite{bonato2022}, while mixtures of semiflexible polymers and rigid threadable rings yield percolating bicomponent networks driven purely by steric repulsion~\cite{bonato2025prl}. Furthermore, recent work has shown that external driving, such as shear, can manipulate topological entanglements to encode mechanical memory and induce structural transitions~\cite{palombo2025}.

Despite these advances, existing research has primarily focused on how topological gels form and how their equilibrium connectivity scales with physical parameters. However, 
much less is known about whether an already self-assembled topological material can be programmed and controlled post-assembly without altering its microscopic constituents or relying on continuous external driving.

Here, we show that such post-assembly control is achievable by exploiting three physical properties of the topological entanglements: mobility, confinement geometry, and dynamical rewiring. By simulating a bicomponent mixture of semiflexible polymers and rigid colloidal rings~\cite{bonato2025prl}, we address the following questions. First, can we control macroscale mechanical response by varying the mobility of topological constraints without changing the microscopic origin of the entanglements? Second, can the geometry of spatial confinement template the architecture of the self-assembled entanglement network?
Finally, can we dynamically rewire the entanglements to reconfigure the network post-assembly?


To answer these questions, we consider the same microscopic system but subjected to different anchoring, confinement, and reconnection protocols. We find that these three problems lead to three different control modes for topological networks post-assembly. 

First, we show that the mobility of topological constraints determines the mechanical response of the network. Thus, quenched anchoring locks topological entanglements in place, giving rise to 
highly stretched load-bearing topologically linked networks which eventually yield. Instead, annealed anchoring allows polymer ends to slide along surfaces, enabling continuous stress relaxation whilst preserving global connectivity of the networks. 
Second, we find that spatially confining boundaries act as geometrical templates that imprints its topology onto the self-assembled material. Thus, spherical confinement produces compact 3D gels, whereas spherical shell confinement templates surface-spanning 2D nets or quasi-1D topological rings, depending on the anchoring protocol. These networks are characterised by distinct graph-spectral signatures, which capture their fundamentally different macroscopic architectures.
Finally, we show that allowing dynamic polymer reconnection, a process analogous to DNA recombination, leads to a global rewiring of the entanglement network which controls topological complexity. In contrast to pure polymer systems, where reconnection generates complex, highly linked and knotted ring networks~\cite{bonato2025prl}, the presence of rigid colloidal rings simplifies topological complexity, restricting it to simple, pairwise Hopf links alongside mechanically trapped, geometrically interlocked states.


Together, these results shift the paradigm of topological soft matter from passive equilibrium formation to active dynamical control. By demonstrating how microscopic entanglement mobility, surface templating, and topological rewiring dictate macroscopic mechanics, our findings provide concrete physical principles for designing adaptive soft metamaterials, synthetic slide-ring networks, and biomimetic DNA assemblies.


\section{Our model: a topological bi-gel}

Unless otherwise specified, our model system consists of a bicomponent mixture comprising $N_p=30$ semiflexible linear polymers and $N_r = 90$ rigid colloidal rings. Each polymer comprises $L_p = 108$ beads of diameter $\sigma$. Consecutive beads along a chain are connected by finitely extensible nonlinear elastic (FENE) bonds, while the chain bending stiffness yields a persistence length of $l_p = 5\sigma$ via a Kratky-Porod potential. The rigid colloidal rings have radius $R_r = 6.4\sigma$, and they are modeled as planar circular arrays consisting of $L_r = \lfloor 2\pi R_r / \sigma \rfloor = 40$ rigid beads of size $\sigma$, where $\lfloor \cdot \rfloor$ denotes the floor function. As established in~\cite{bonato2025prl}, this parameter set places the system inside the topological gel phase.

\begin{figure}[t!]
	\includegraphics[width=0.9\textwidth]{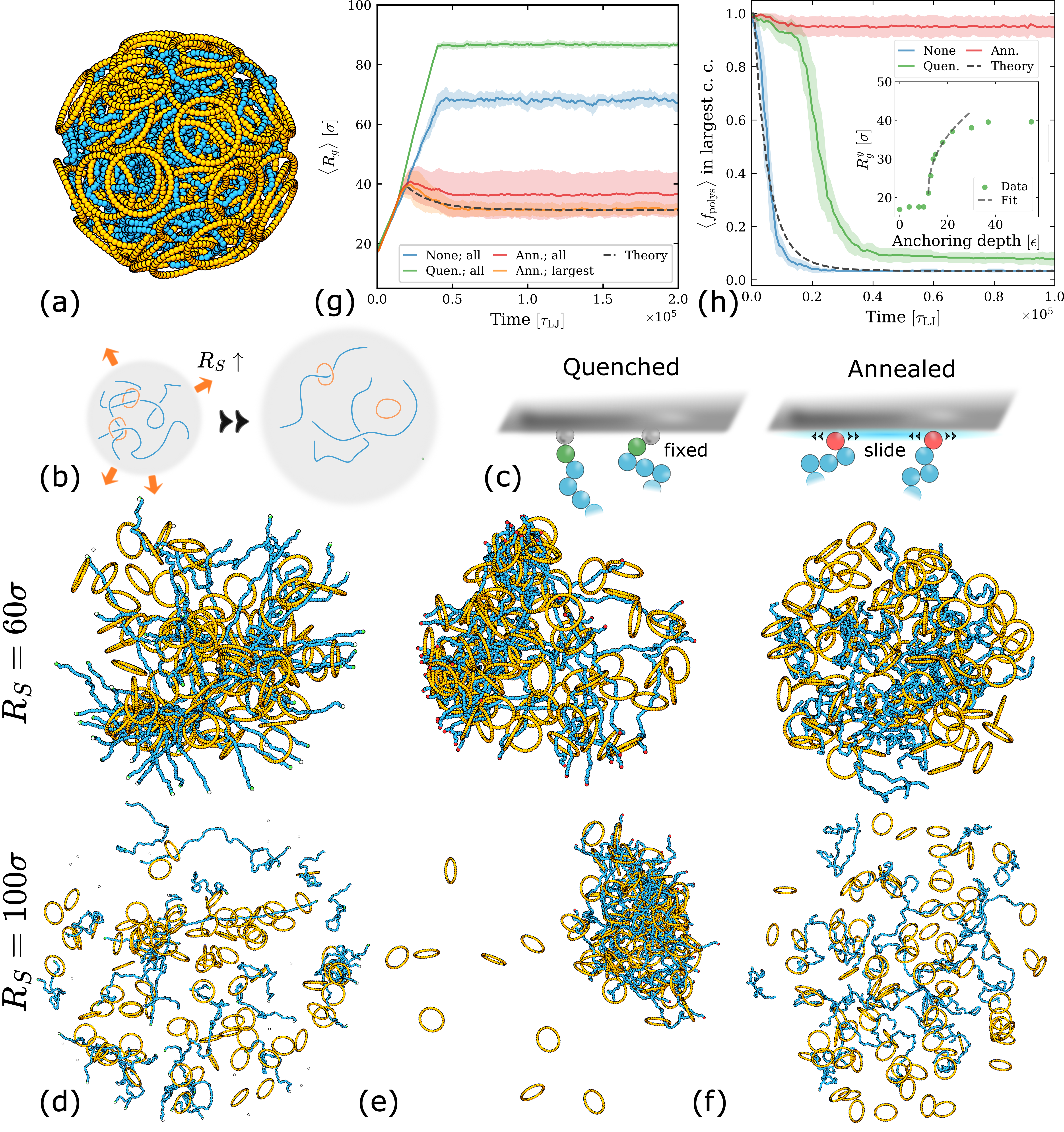}
    \caption{{\bf Spontaneous bulk expansion.} 
 Simulation of bi-gels within an expanding confinement sphere without anchoring and with quenched and annealed anchoring. \textbf{(a)} Snapshots taken at initial sphere radius $R_S=R_0=25\sigma$ without anchoring. \textbf{(b)} Sketch of radial expansion with no anchoring. \textbf{(c)} Sketch of quenched anchoring and annealed anchoring.\textbf{(d) (e) (f)} Snapshots taken at sphere radius $R_S=60\sigma$ and $R_S=100\sigma$ with quenched anchoring \textbf{(d)}, annealed anchoring \textbf{(e)} and without anchoring \textbf{(f)}. \textbf{(g)} Radius of gyration of all the polymers (all) and for the polymers belonging to the largest connected component (largest) for different anchoring styles as a function of time. \textbf{(h)} Fraction of polymers in the largest connected component for different anchoring styles as a function of time. The inset shows the yielding elongation $R_g^y$ as a function of anchoring potential depth. Quen.: quenched; ann.: annealed; polys: polymers; c. c.: connected component.}
\label{fig_anch}
\end{figure}

All bead-bead interactions occur through a purely repulsive Weeks-Chandler-Anderson potential. Molecular dynamics simulations are executed using LAMMPS~\cite{LAMMPS}, employing a Langevin thermostat and a time step of 
$\Delta t= 0.01\,\tau_{LJ}$. Here $\tau_{LJ}=\sqrt{m\sigma^2/\epsilon}$ represents the Lennard-Jones time unit, where $m$ is the mass of the beads and $\epsilon=1k_BT$ is the energy unit.
The mixture is initially prepared within a large confining sphere that is gradually compressed and equilibrated to reach the desired radius $R_S=R_0=25\sigma$. This setup (Fig.~\ref{fig_anch}a) results in an initial volume fraction $\phi=(N_pL_p+N_rL_r)\sigma^3/(8R_0^3) \sim  0.05$. This equilibrated gel configuration phase serves as the baseline for all subsequent dynamic processes explored in this study (see Supplemental Material for full simulation details).

\section{Results}


\subsection{Constraint mobility switches the gel from rigid to adaptive}

A defining feature of the bi-gel is that its macroscopic network connectivity coexists with continuous microscopic topological rearrangement. Rings thread and unthread polymers on timescales much shorter than the lifetime of the percolating network, so the material is connected but its individual constraints remain mobile~\cite{bonato2025prl}.  We shall exploit this separation between global architecture and local constraint dynamics to programme the network post-assembly. 

We first study the response of the bi-gel by expanding the spherical confinement. Upon expansion, a bi-gel with unanchored (free) polymer ends dilutes, causing ring-polymer threadings to unbind until the percolating cluster progressively dissolves. Anchoring the polymer ends prevents network disassembly; however, the resulting mechanical response depends critically on whether the anchor points are fixed in space (\emph{quenched anchoring}) or free to slide along the boundary surface (\emph{annealed anchoring}).

As shown in Fig.~\ref{fig_anch}, the corresponding three protocols lead to fundamentally different outcomes qualitatively. With free ends, topological entanglements escape and the network dissolves (Fig.~\ref{fig_anch}f). With quenched anchoring (Fig.~\ref{fig_anch}c left), the entanglements cannot redistribute (Fig.~\ref{fig_anch}d). As a consequence the polymers stretch, and elastic stress is stored until the network yields. With annealed anchoring (Fig.~\ref{fig_anch}c right), the polymer ends, by sliding along the surface, can redistribute the topological load. Following an initial deformation, the network dynamically contracts and maintains almost all of its global topological connectivity (Fig.~\ref{fig_anch}e). Thus, identical microscopic types of topological interactions can produce either stress storage or stress relaxation, depending on the anchoring protocol considered (which controls the mobility of the topological constraints on the surface).

The elastic response of these systems during expansion is quantified in Fig.~\ref {fig_anch}g,  where we report the time evolution of the average radius of gyration of the polymers, $R_g(t)$. For quenched anchoring, the stretching of the polymers is highlighted by the rapid growth of $R_g$, which plateaus after network rupture ($t \approx 5 \times 10^4\,\tau_B$) at a value higher than in the unanchored case due to the residual anchoring of some of the ends to the encasing sphere. In contrast, annealed anchoring exhibits a non-monotonic elastic response over time: an initial increase in $R_g$, driven by partial stretching and consistent with the two previous cases, is followed by a subsequent contraction. This contraction occurs as the polymer ends move closer along the surface, relaxing the accumulated stress.

The relaxation dynamics of the bi-gel with annealed anchoring can be modelled by treating the largest topologically connected component as an effective entropic spring of end-to-end extent $2R_g$, whose boundary terminals undergo overdamped motion on the confining surface. The relaxation toward the equilibrium size $R_g^{eq}$ is governed by the following equation,
\begin{equation}
R_g(t) = R_g^{eq}+(R^0_g-R_g^{eq})\,e^{-(t-t_0)k_R/\zeta_S},
\label{eq:relax_ann}
\end{equation}
where $k_R$ is the effective network spring constant, $t_0$ marks the onset of relaxation, and $\zeta_S$ is the effective surface friction experienced by the sliding ends.
Eq.~\eqref{eq:relax_ann} provides an excellent fit to the annealed system data in Fig.~\ref{fig_anch}g, with $t_0 = 2 \times 10^4\,\tau_B$, $R_{\text{g}}^0 = 39\sigma$, $R_{\text{g}}^{\text{eq}} = 31.35 \pm 0.05\,\sigma$, and relaxation rate $k_R / \zeta_S = (5.6 \pm 0.2) \times 10^{-5}\,\tau_B^{-1}$. The parameter $\zeta_S$ characterises the boundary mobility and provides a dial to tune the elastic response of the system based on annealed anchoring.

We next examine the evolution of network topology, tracked by $f_{\text{max}}(t)$, the fraction of polymers belonging to the largest topologically connected cluster (Fig.~\ref{fig_anch}h).

In the absence of anchoring, expansion causes rapid network dissolution. For slow expansion, this decay can be captured by the following simple argument. As established in Ref.~\cite{bonato2025prl}, $f_{\text{max}}$ depends on the single-threading probability $q$ according to:
\begin{equation}
f_{\text{max}}(t) = \min \left(1, \, \frac{N_r}{N_p} \sum_{n_{\text{th}} \ge 2} \pi(n_{\text{th}})(n_{\text{th}} - 1) + \frac{1}{N_p} \right),
\label{frmax}
\end{equation}
where $n_{\text{th}}$ denotes the number of threadings, and where the threading distribution $\pi(x)$ follows binomial statistics:
\begin{equation}
\pi(x) = \binom{N_p}{x} q^x (1 - q)^{N_p - x}.
\label{bin}
\end{equation}
During expansion, the threading probability depends on time, $q=q(t)$. Its time dependence can be estimated by noting that, within the parameter space investigated (at fixed $\phi=0.05$), Ref.~\cite{bonato2025prl} supported a linear growth of $q$ with the $R_{g}$ of both components, and simple dimensional analysis suggests that $q \propto R_S^{-2}$, where $R_{S}$ is the radius of the confining sphere. On the other hand, in our expansion protocol $R_S=R_0+rt$, where $r=1.875\cdot 10^{-3}$ $\sigma\,\tau^{-1}_{B}$ is the expansion rate, and  $R_0=25\sigma$ is the initial radius of the confinement sphere. Hence, by assuming $q(t) = q_0(R_0+rt)^{-2}$, the functional form in Eq.~\eqref{frmax} combined with Eq.~\eqref{bin} fits the evolution in time of the fraction of polymers in the largest connected component of the gel without anchoring reported in Fig.~\ref{fig_anch}h, providing the estimate  $q_0=21.3\pm0.4\,\sigma^2$. Note that, by assuming the more general dependence $q(t) = q_0(R_0+rt)^{-\alpha}$ and performing a two-parameter fit, we get $\alpha=1.9\pm0.1$  and $q_0=16\pm6\,\sigma^2$, which are consistent with the quadratic dependence.

The bi-gel with quenched anchoring presents instead a two-stage topological response (Fig.~\ref{fig_anch}h). At early times, polymer stretching dominates and $f_{\text{max}}$ decreases linearly as isolated threadings unbind. Beyond a critical yield elongation $R_g^y$, stored elastic tension exceeds the anchoring energy, triggering much faster network dissolution. These two regimes can also be observed in the network representations in Fig.~\ref{fig_anch_net_SI}, which show how the largest cluster, first encompassing all polymers in the system, progressively disassembles, but, for an initial ``yielding'' time interval, it does so at a much slower rate.
Investigating the behavior of $R_{g}^{y}$ as a function of the anchoring potential depth $M_{\epsilon }$ (inset of Fig.~\ref{fig_anch}h) reveals distinct anchoring regimes. For weak anchoring ($M_{\epsilon}\leq M^c_{\epsilon}\approx 12\,\epsilon$), the surface potential is insufficient to counter the thermal and entropic fluctuations of the system, failing to bind the gel to the surface. Consequently, the yielding regime disappears (see also SI, Fig.~S1) and $R_g^y=R_g(t=t_0)$. Conversely, for high anchoring strengths ($M_{\epsilon}\gtrsim22\,\epsilon$), $R_{g}^{y}$ approaches a plateau; this saturation arises from the finite extensibility of the polymer bonds (see SI), which restricts further deformation as the network reaches its stretching limit. For intermediate anchoring strengths ($12\,\epsilon\lesssim M_{\epsilon}\lesssim22\,\epsilon$), $R_{g}^{y}$ is well-described by the power law $R_g^y \sim (M_{\epsilon}-M^c_{\epsilon})^\alpha$, with $\alpha = 0.44\pm0.05$.
This exponent, which is close to 0.5, indicates that the gel behaves like an entropic spring. Under this framework, the yielding threshold $R_{\text{g}}^{y}$ occurs when the accumulated elastic energy equals the total anchoring detachment energy $E_{\text{anch}}$: \begin{equation}
\frac{1}{2} k_R \left(2 R_{\text{g}}^y - 2 R_{\text{g}}^{\text{eq}}\right)^2 = E_{\text{anch}},
\label{rupture}
\end{equation}
 where $E_{\text{anch}} \approx 2 N_p \Delta U_{\text{Morse}}$ represents the depth of the surface attraction potential for the $2N_{p}$ polymer ends.


Note that, assuming that $R^{eq}_g$ is similar 
for the annealed and quenched gels, combining Eq. \eqref{rupture} with Eq. \eqref{eq:relax_ann} would provide a way to estimate the surface friction of the annealed gel as a function of the yielding radius of the quenched gel from the data.

Finally, if the expansion of the bi-gel occurs with annealed anchoring, the outcome is completely different. The sliding-assisted elastic contraction observed in Fig.~\ref{fig_anch}g relieves chain tension, keeping the topological entanglement network intact. 
This is shown by the fact that $f_{\text{max}}(t)$ remains nearly constant throughout expansion, demonstrating that mobile topological entanglements convert an otherwise brittle gel into a resilient, stress-relaxing adaptive material (see Fig.~\ref{fig_anch}h and  Fig.~\ref{fig_anch_net_SI}).

\subsection{Mobile topological constraints enable sustained mechanical expansion}

\begin{figure}[t!]
	\includegraphics[width=0.8\textwidth]{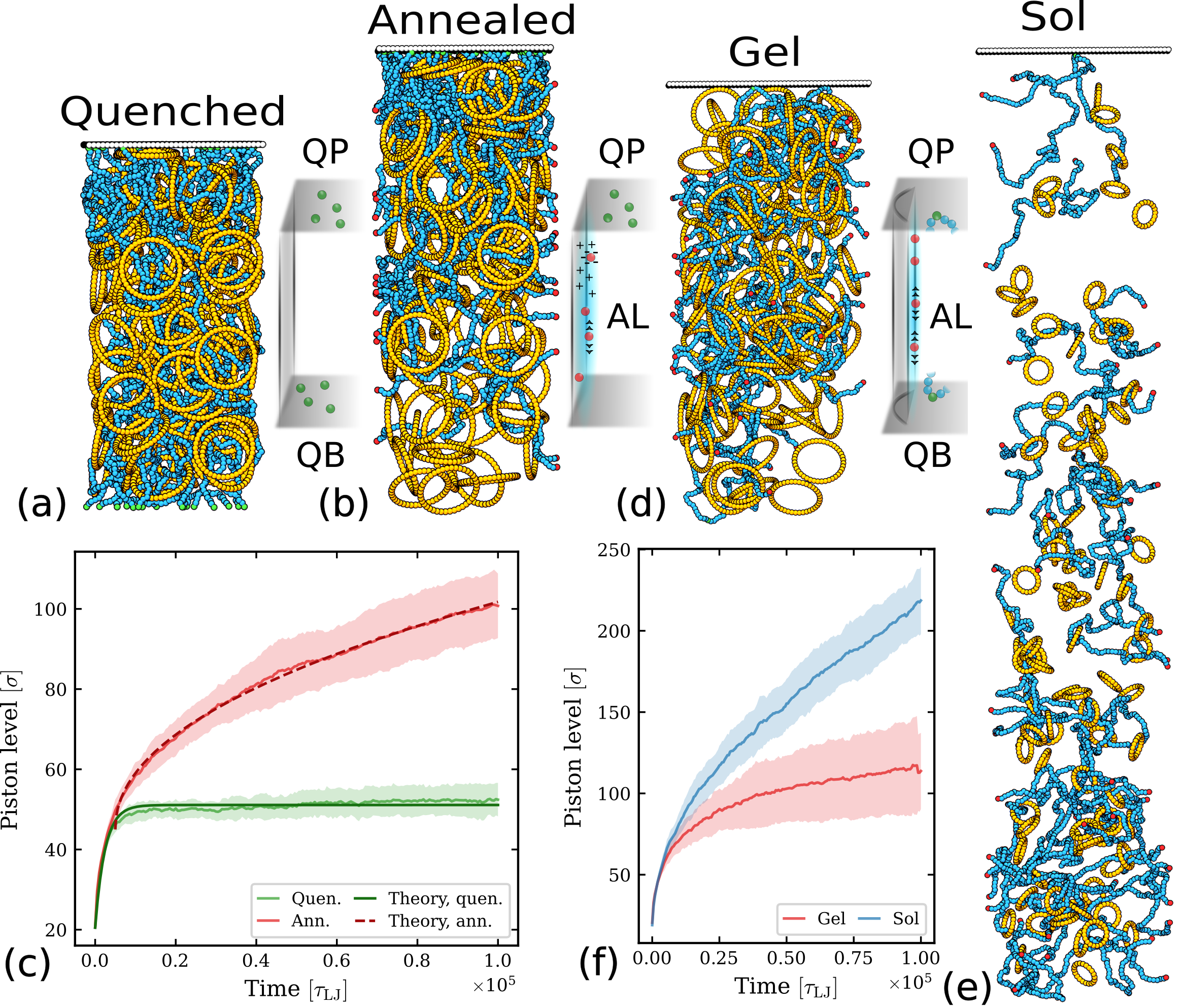}
    \caption{{\bf Elastic response.} 
    Simulations of anchored bi-gels within a box with a movable wall. \textbf{(a) (b)} Snapshots from simulations with top and bottom quenched anchoring (QP-QB). \textbf{(a)} and from simulations with quenched anchoring on top (QP) combined with annealed anchoring on the lateral walls (AL). \textbf{(b)} Sketches of the anchoring protocols are depicted to the right of the snapshots. \textbf{(c)} Height of the piston with QB-QP (quen., green) and AL-QP (ann., red) protocols as a function of time. \textbf{(d) (e)} Snapshots from simulations of the gel \textbf{(d)} and sol \textbf{(e)} phases with QP-AL-QB anchoring, where just one polymer is anchored to the top, and one to the bottom and bottom (see sketch in the middle). \textbf{(f)} Piston level for the QP-AL-QB protocol as a function of time. Quen.: quenched; ann.: annealed.
    }
\label{fig_piston}
\end{figure}

While the spherical expansion protocol just considered demonstrates that constraint mobility governs internal stress relaxation, harnessing this property for mechanical function requires converting internal tension into macroscale mechanical work. To test this, we place the topological bi-gel in a piston geometry, inspired by experiments on biological systems~\cite{pelletier2012}, and compare two canonical boundary conditions: a \emph{locked geometry} and a \emph{sliding geometry}. In the locked geometry (\emph{QP-QB}), polymer ends are quenched (fixed) both at the movable piston head and at the fixed base (Fig.~\ref{fig_piston}a). In the sliding geometry (\emph{AL-QP}), polymer terminals on the piston remain quenched, whereas all other boundary anchors are annealed, allowing polymer ends to slide along the lateral walls (Fig.~\ref{fig_piston}b).


Measuring the time-dependent piston displacement $z(t)$ reveals fundamentally distinct mechanical regimes (Fig.~\ref{fig_piston}c). Although both systems expand initially, the piston level in the locked geometry quickly saturates to a stationary height. In contrast, the sliding geometry exhibits sustained, unbounded upward motion, dragging the entangled bi-gel along the container walls. 

The latter result shows how boundary mobility dictates bulk mechanics. A locked bi-gel behaves as a conventional elastic material in which the piston approaches a stationary height determined by the balance between the osmotic driving and the elastic restoring force. A sliding bi-gel behaves instead as an adaptive fluid-like material. This is because the lateral anchor mobility enables internal threadings and entanglements to redistribute tangentially along polymer backbones, which suppresses stress accumulation and instead enables sustained piston motion, which generates work.

We model the locked piston dynamics by treating the bulk network as an effective entropic spring with spring constant $k_z$ experiencing linear viscous damping with an effective drag coefficient $\zeta_z$. Overdamped force balance yields:
\begin{equation}
\zeta_z \frac{dz}{dt} = -k_z (z - z_{\text{eq}}),
\end{equation}
where $z_{\text{eq}}$ is the equilibrium piston height under load. Integrating from an initial compressed state $z_0$ gives exponential relaxation toward saturation:
\begin{equation}
z(t) = z_{\text{eq}} - (z_{\text{eq}} - z_0) e^{-(k_z / \zeta_z) t}.
\label{eq:piston_spring}
\end{equation}
Fitting Eq.~\eqref{eq:piston_spring} to the data reported in Fig.~\ref{fig_piston}c gives $z_{\text{eq}}=(51.1\pm0.08)\,\sigma$ and $k_z/\zeta_z = (4.1 \pm 0.01)\cdot 10^{-4} \tau^{-1}_{LJ}$.
Note that, since the system is isotropic, $k_x=k_y=k_z$; namely, we would have obtained the same relaxation behaviour if we applied the same protocol in the other directions.

The sliding geometry case is different. After the network stretches to its initial relaxation threshold at time $t_0$, sustained piston motion is propelled by a population of $N_{\text{spare}}$ unentangled rigid rings that phase-separate to the base of the chamber (Fig.~\ref{fig_piston}b).
Treating these unthreaded rings as an ideal gas exerting an osmotic pressure $p = N_{\text{spare}} k_B T / V$ against an effective piston friction factor $\zeta_{\text{piston}}$
, which depends on the properties of the piston as well as the polymeric material anchoring it to the lateral walls, the hydrodynamic force balance $p V = \zeta_{\text{piston}} (z - z_{\text{eq}}) (dz/dt)$ yields:
\begin{equation}
(z - z_{\text{eq}}) \frac{dz}{dt} = \frac{N_{\text{spare}} k_B T}{\zeta_{\text{piston}}}.
\end{equation}
Integrating yields a diffusive, square-root expansion power law:
\begin{equation}
z(t) = z_{\text{eq}} + \sqrt{\frac{2 N_{\text{spare}} k_B T}{\zeta_{\text{piston}}} (t - t_0)}, \quad \text{for } t > t_0.
\label{eq:gas}
\end{equation}
Therefore, lateral sliding transforms the exponential approach to saturation seen in locked gels into a sustained $z(t) \propto t^{1/2}$ power-law expansion. 
Fitting the sliding geometry trajectories to $z(t) = z_{eq}+a(t-t_0)^{\alpha}$ with $t_0=5\cdot10^3\tau_{LJ}$ we get $z_{eq}=45.0\pm0.4$, $\alpha=0.474\pm0.007$, values compatible with the predictions (see Fig.~\ref{fig_piston}c). 

To demonstrate how global network percolation dictates this mechanical response, we compare the expansion dynamics of the percolating bi-gel phase against a non-percolating sol phase ($R_r = 3.2\sigma$) under identical anchoring conditions (Fig.~\ref{fig_piston}d-f and SI, Fig.~S5). Under fully locked conditions (QP-QB; SI, Fig.~S5b), the sol expands slightly less than the gel due to lower entropic ring pressure. 
Conversely, when annealed anchoring is allowed at the lateral walls (AL-QP; SI, Fig.~S5a), the situation reverses, with the gel phase exhibiting a slower piston rise rate compared to its sol counterpart. This behaviour is expected, given that the phase-separated gas of rings contains a higher particle count in the sol phase (higher $N_{spare}$), and that the effective friction $\zeta_{\text{piston}}$ should be higher in the gel phase. 
In this regard, we point out that, to keep the same initial volume fraction, the sol system was confined within a smaller cross-sectional area. 

The contrast between gel and sol mechanics becomes most pronounced when testing topological load-bearing capacity directly: we anchor just \emph{two} individual polymer chains—one to the moving piston head and one to the fixed container base—while allowing all remaining chain ends to slide along the lateral walls (QP-AL-QB; Fig.~\ref{fig_piston}d,e). In the percolating gel phase, a single continuous path of interlocked rings and polymers connects the two anchored chains, effectively ``welding'' the piston to the base and halting mechanical expansion (Fig.~\ref{fig_piston}f). In contrast, two randomly chosen chains in the sol phase rarely belong to the same linked component; lacking a continuous topological pathway across the cylinder, the sol system cannot support tensile stress, allowing the unthreaded rings to drive unhindered piston motion. Thus, a single percolating topological path suffices to lock macroscopic expansion, providing a highly sensitive topological switch for mechanical actuation.

\subsection{Chemical control of connectivity generates mechanical work}

\begin{figure}[t!]
	\includegraphics[width=0.65\textwidth]{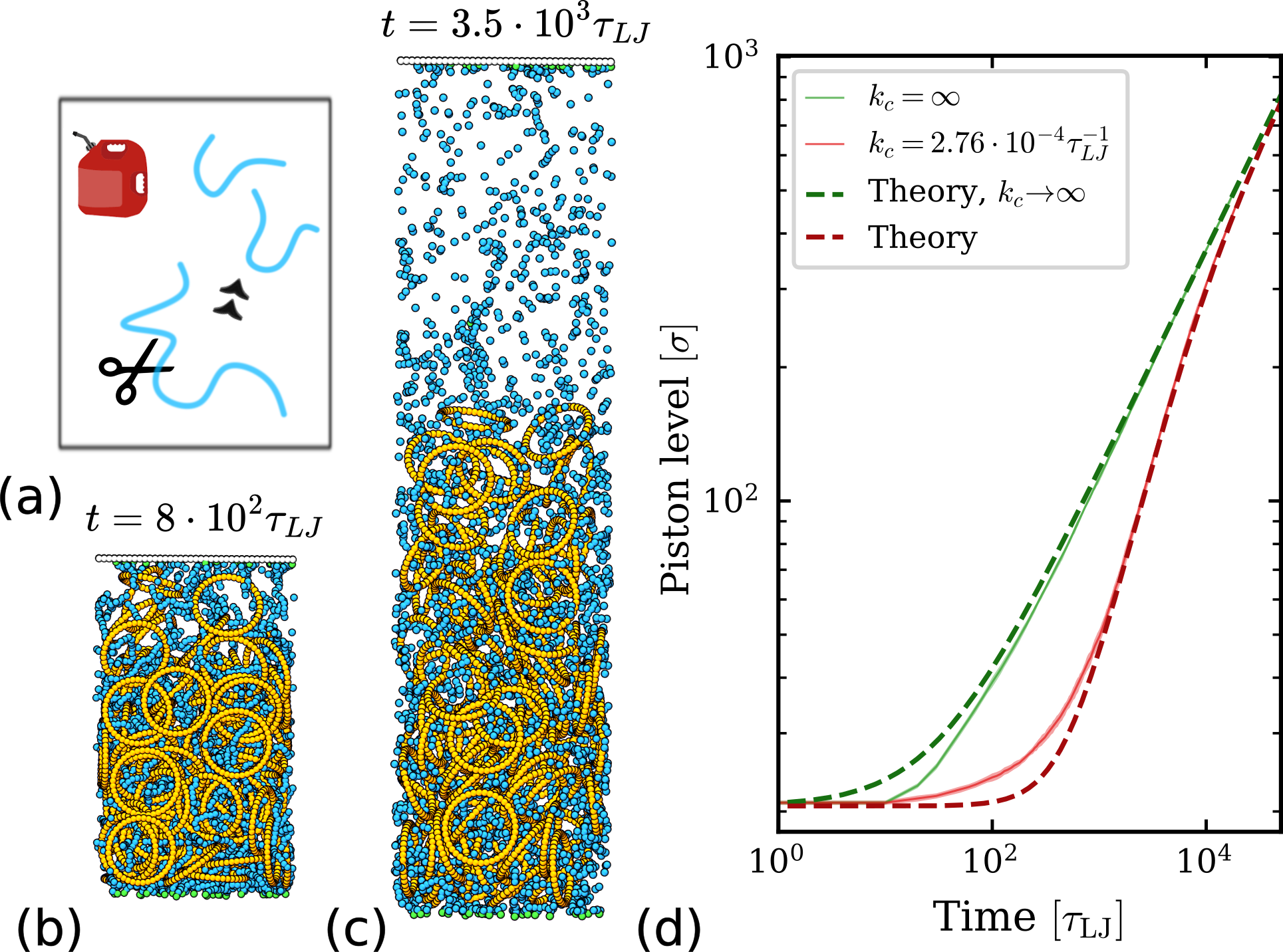}
    \caption{{\bf Expansion dynamics with bond cleavage reaction.} Results from simulations of a gel subject to a bond cleavage chemical reaction expanding within a piston. \textbf{(a)} Sketch of the chemical reaction.
    \textbf{(b) (c)} Snapshots taken at two different times, $t=8\cdot 10^2\tau_{LJ}$ and $t=3.5\cdot 10^3\tau_{LJ}$, with reaction rate $k_c=2.76\cdot10^{-4}\tau_{LJ}^{-1}$.  \textbf{(d)} Piston level as a function of time for two reaction rates, $k_c=2.76\cdot10^{-4}\tau_{LJ}^{-1}$ and $k_c=\infty$.
    }
\label{fig_piston_rocket}
\end{figure}

The preceding results establish that adjusting the boundary mobility of topological entanglements controls stress relaxation and mechanical work. A complementary route to actuation is to dynamically alter the density of mechanically active constituents via chemical reactions. To illustrate this possibility, we introduce bond cleavage at a controlled rate $k_c=1/\tau_c$, as sketched in Fig.~\ref{fig_piston_rocket}. Polymer cleavage progressively breaks the chains into shorter, unentangled fragments, simultaneously releasing topological constraints and increasing the effective particle number from $N_p$ chains to $N_b = N_p L_p$ individual monomers (Fig.~\ref{fig_piston_rocket}b,c). This chemical fragmentation converts a topologically connected gel into an expanding gas, generating osmotic power.


We model this process by treating the cleaving mixture as an ideal gas with a time-dependent particle number $N(t) = N_b (1 - e^{-t/\tau_c})$. Balancing the time-dependent osmotic pressure $p(t) = N(t) k_B T / V(t)$ against the effective piston friction $\zeta_{\text{piston}}$ (which we assume to be time-independent, after an initial transient) yields the hydrodynamic equation of motion:
\begin{equation}
\zeta_{\text{piston}}z(t) \frac{dz}{dt} = N_b k_B T \left(1 - e^{-t/\tau_c}\right).
\end{equation}
Integrating from an initial piston position $z(0) = z_0$ yields the trajectory for chemomechanical expansion as
\begin{equation}
z(t) = \sqrt{\frac{2 N_b k_B T}{\zeta_{\text{piston}}} \left( t + \tau_c e^{-t/\tau_c} - \tau_c \right) + z_0^2}.
\label{eq:rocket}
\end{equation}
Eq.~\eqref{eq:rocket} displays two distinct dynamical regimes: at early times ($t \ll \tau_c$), Taylor expansion of the exponential term yields an initial constant-velocity phase, $z(t) \approx \sqrt{z_0^2 + (N_b k_B T / \zeta_{\text{piston}} \tau_c) t^2}$; at long times ($t \gg \tau_c$), complete fragmentation transforms the dynamics into the characteristic diffusive power law, $z(t) \sim \sqrt{t}$. As shown in Fig.~\ref{fig_piston_rocket}d, $z(t)=\sqrt{ at+z_0^2}$ fits our molecular dynamics simulations with $\tau_c=0$ very well with $a=(13.451\pm0.004)\sigma^2/\tau_{LJ}$, and Eq.~\eqref{eq:rocket} with $\frac{2N_bk_BT}{\zeta_{\text{piston}}}=a=13.451\sigma^2/\tau_{LJ}$ closely matches our simulations with $k_c=\tau_c^{-1}=2.76\cdot10^{-4}\tau^{-1}_{LJ}$, confirming that chemical control of chain connectivity offers a programmable mechanism for topological actuation to generate mechanical work.


\subsection{Confinement geometry templates network architecture and graph spectra}

\begin{figure}[t!]
	\includegraphics[width=0.95\textwidth]{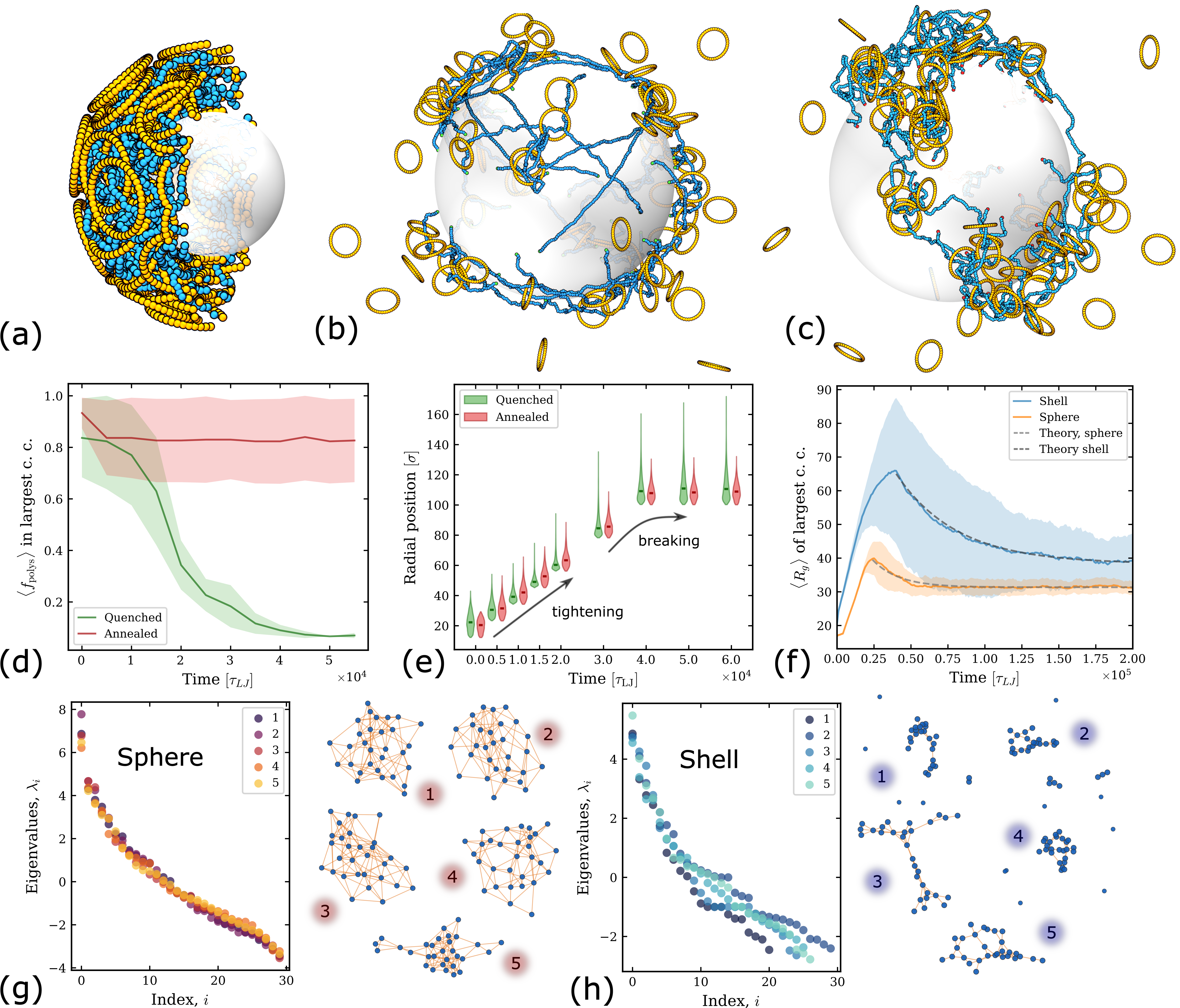}
    \caption{{\bf Spherical shell confinement.} 
    Simulations of anchored bi-gels within an expanding confinement shell. \textbf{(a)} Snapshot from simulations of a bi-gel assembled within a spherical shell (before anchoring).  \textbf{(b)} Snapshot taken at inner sphere radius $R_S=40\sigma$ with quenched anchoring.
    \textbf{(c)} Snapshot taken at inner sphere radius $R_S=40\sigma$ with annealed anchoring.
    \textbf{(d)} Fraction of polymers in the largest connected component as a function of time.
    \textbf{(e)} Violin plot of the radial position of all polymer beads as a function of time.
    \textbf{(f)} $\langle R_g \rangle$ of the largest connected component as a function of time for a gel assembled within a sphere (orange) and within a spherical shell (blue) with annealed anchoring.
    \textbf{(g) (h)} Spectra of the adjacency matrices of 10 different structures assembled with annealed anchoring, half of which are within a sphere \textbf{(g)} and the other half within a shell \textbf{(h)}. Each spectrum is accompanied by the network representation of the corresponding structure. Polys: polymers; c. c.: connected component.
    }
\label{fig_shell}
\end{figure}

While the preceding sections manipulated boundary mobility in a fixed geometry, in this Section we now ask whether spatial geometry can be used to control the macroscopic architecture and elastic response of self-assembled gel. To this end, we compare self-assembly in a three-dimensional spherical cavity with assembly within a spherical shell (see Supplemental Material for more details on gel preparation). The microscopic constituents, concentration and interaction rules are unchanged. Yet, the resulting architectures and topological spectra are qualitatively very different. Spherical confinement produces compact, isotropic, and highly interconnected 3D topological gels.  In contrast, confinement within a spherical shell forces the network into lower-dimensional patterns. 
Quenched anchoring on a shell generates a quasi-2D surface-spanning topological net wrapping the inner sphere (Fig.~\ref{fig_shell}b). Instead, annealed anchoring allows lateral sliding and, as a result, while relaxing, the network reorganises into a quasi-1D macroscopic branched or cyclical assembly interlocked by colloidal rings (Fig.~\ref{fig_shell}c, and SI Fig.~S4).


The elasticity of shell-generated networks under radial expansion can be studied by letting the radius of the anchored inner surface grow over time following the removal of the outer boundary. The observed response mirrors the mobility-dependent dynamics observed for spherical confinement (Fig.~\ref{fig_shell}d--f). Under quenched anchoring, radial expansion stretches the surface network, causing chains to tighten tightly against the shell, as evidenced by a narrowing radial position distribution of polymer beads (Fig.~\ref{fig_shell}e). Beyond a critical expansion threshold, boundary tethers detach, causing catastrophic network tearing and a rapid drop in the giant component fraction $f_{\text{max}}(t)$ (Fig.~\ref{fig_shell}d). Conversely, under annealed anchoring, polymers slide laterally along the surface, preserving global cluster connectivity ($f_{\text{max}} \approx 1$) (Fig.~\ref{fig_shell}d,e).

To quantify the elasticity of the different bi-gels, we fit the entropic relaxation model (Eq.~\eqref{eq:relax_ann}) to the time evolution of the radius of gyration $R_g(t)$ following expansion (Fig.~\ref{fig_shell}f). Fitting yields relaxation rates $b_{Sh} = k_{Sh}/\zeta_S = (2.44\pm0.07)\cdot 10^{-5}\tau^{-1}_{LJ}$ for the shell and $b_{Sp} = k_{Sp}/\zeta_S = (5.6 \pm 0.2)\cdot10^{-5} \tau^{-1}_{LJ}$ for the sphere. This results in a stiffness ratio of $k_{Sp}/k_{Sh} \sim 2.3$, which is significantly higher than the measured ratio of their average total number of links (edges in the graph representation), $\langle N^{Sp}_{bn}\rangle/\langle N^{Sh}_{bn}\rangle\sim1.4$ (the number of bonds is obtained from the adjacency matrix of a graph $A_{ij}$: $N_{bn}=\frac{1}{2}\sum_{ij}A_{ij}$). This discrepancy recapitulates how the anisotropicity of the super-structure generated within a shell deeply impacts its elasticity. In particular, as highlighted above, the annealed shell gel frequently self-assembles into a macroscopic ring; consequently, the relaxation of $R_{g}$ occurs primarily along the radial direction, where network bonds are highly depleted, drastically softening the macroscale network response.

Finally, to quantitatively compare the global topological architectures formed in 3D spheres versus 2D shells, we map each self-assembled bi-gel onto an unweighted topological graph $\mathcal{G} = (\mathcal{V}, \mathcal{E})$. Nodes $i, j \in \mathcal{V}$ represent individual polymer chains and an edge $(i, j) \in \mathcal{E}$ is assigned if both chains simultaneously thread at least one common rigid ring (adjacency matrix $A_{ij} = 1$). 
The resulting graphs, shown in Figs.~\ref{fig_shell}g and h, highlight a stark structural contrast. Even though both gels are assembled at the same volume fraction $\phi $, and despite the largest component spanning the vast majority of the system in both cases, their topologies differ significantly: a gel assembled within a sphere is characterised by a denser, disordered, and isotropic network of bonds; conversely, a gel assembled on a shell exhibits a more brittle, branched, and cyclic structure, with the graphs frequently closing into macroscopic loops.

As shown in Fig.~\ref{fig_shell}g,h, the spectrum of adjacency eigenvalues $\lambda_k$ (sorted in descending order) clearly distinguishes 3D bulk gels from 2D shell assemblies.
The spectrum of the gel assembled inside a sphere (Fig.~\ref{fig_shell}g) is distinguished by a sharp gap between the largest eigenvalue and the rest, followed by an almost linear decay when plotted against the eigenvalue index. This spectral signature reflects a densely connected, isotropic network. Instead, the spectrum of the gel on a spherical shell (Fig.~\ref{fig_shell}h) displays a gapless and more gradual transition from the leading eigenvalue, reflecting a more amorphous network of entanglements. This result shows that graph representation and spectral analysis is a powerful framework to quantitatively capture the structural properties of topological gels.


\begin{figure}[t!]
 \includegraphics[width=0.9\textwidth]{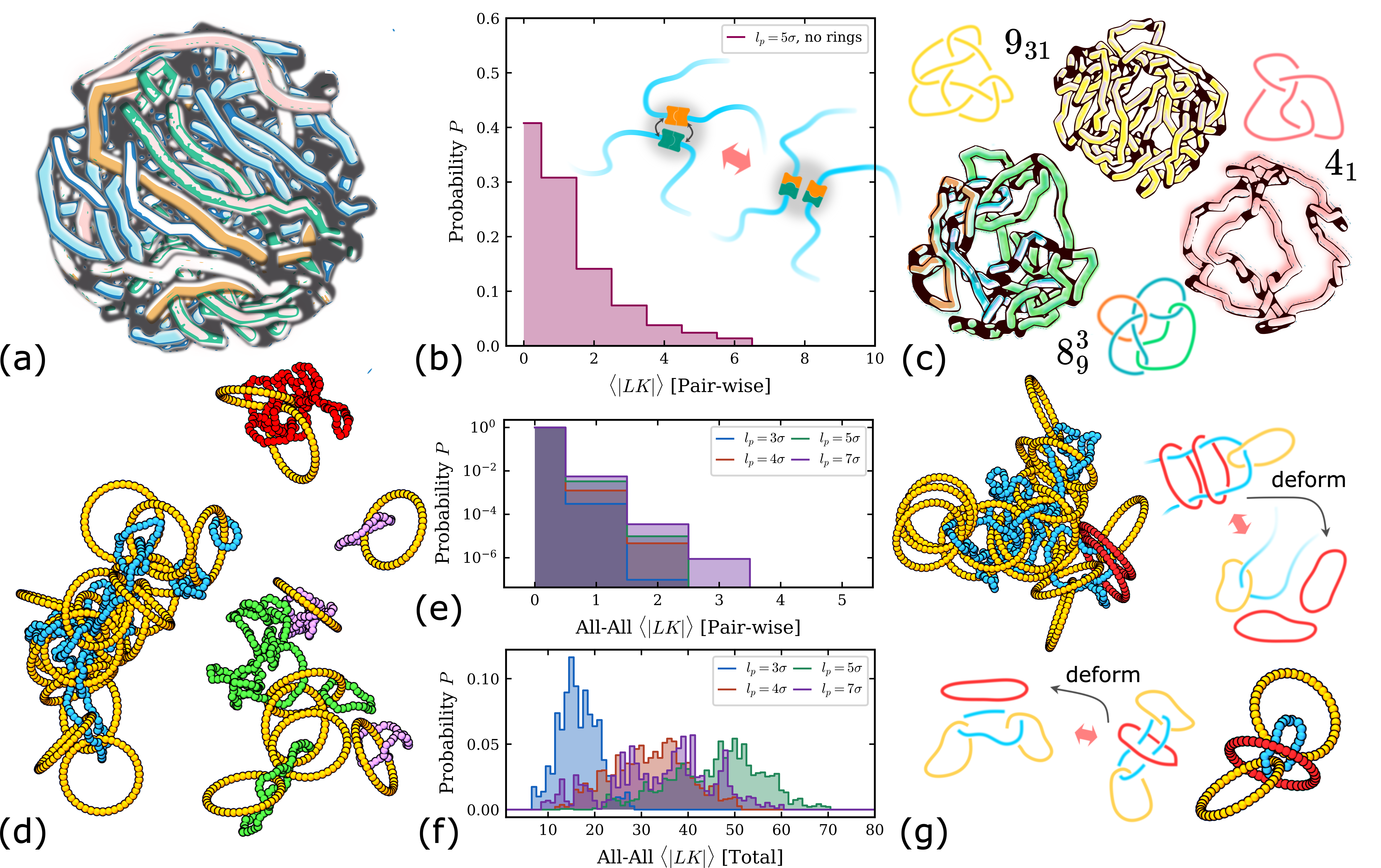}
    \caption{{\bf Topologies in recombined bi-gels and gels.} 
    \textbf{(a-c)} Results from simulations of reconnecting polymers with persistence length $l_p=5\sigma$ under confinement (without rigid rings). \textbf{(a)} Snapshot from simulations. \textbf{(b)} Distribution of the unsigned pairwise linking number. 
    The inset depicts a sketch of a reconnection operation. \textbf{(c)} Example of links and knots contained within the gel. \textbf{(d-g)} Results from simulations of recombined bi-gels. \textbf{(d)} Snapshots from simulations ($l_p=5$ for the polymers). Only linked rings are shown and the three biggest connected components are highlighted with different colours. \textbf{(e)} Distributions of the unsigned all to all pairwise linking number for different persistence lengths. Links can be established between either two ring polymers or a rigid ring and a ring polymers. Here we do not distinguish between ring polymers and rigid rings, hence the denomination all to all (all-all). \textbf{(f)} Distribution of the total all to all linking number (sum of the corresponding unsigned pairwise all-all quantity). \textbf{(g)} Example of links within the recombined bi-gel. Red rings are topologically unlinked under isotopies, but linked under isometries.
    }
\label{fig_topo}
\end{figure}

\subsection{Reconnection rewrites the assembled topology}

While boundary mobility and confinement manipulate pre-existing topological constraints within a fixed network topology, strand reconnection introduces an active rewrite operation that fundamentally alters component connectivity. In biological and synthetic polymer systems, cut-and-paste reconnection (e.g., topoisomerase-mediated cleavage, reversible ligation, or reversible polymerisation in melts of living polymers~\cite{Cates1990,Petschek1986}), transforms open linear chains into closed ring topologies~\cite{Alberts2014}. For the results in this Section, we permit equilibrium strand reconnection among the flexible polymers within the pre-assembled bi-gel and analyse how the presence of the rigid-ring scaffold alters the resulting topological landscape (Fig.~\ref{fig_topo}).

As highlighted in Fig.~\ref{fig_topo}a-c, a distinctive characteristic of gels created by reconnected polymers (see~\cite{bonato2022}) is the highly intricate linking and knotting topology of their components. Interestingly, introducing a scaffold of rigid rings, as in the bi-gel, drastically alters the topological landscape. Fig.~\ref{fig_topo}d-f illustrates that, upon introducing reconnection in the bi-gel, the linear polymers transform into ring polymers, effectively trapping the rigid rings within clusters comprising multiple components, with the number of connections depending on the persistence length $l_p$ (Fig.~\ref{fig_topo}f). The key insight, as depicted in Fig.~\ref{fig_topo}e, is the observation that the pairwise linking number of the linked system's components sharply peaks at 1. This indicates that any two constituents are either unlinked or Hopf-linked. This is a consequence of the fact that the baseline topological interaction between polymers and rings is threading.

A remarkable and unique topological feature of the mixture of rigid and flexible rings is highlighted in Fig.~\ref{fig_topo}g. Given that, according to standard knot theory, two links are topologically equivalent if they can be deformed into each other, if the clusters in the snapshots are interpreted as standard soft links, then the red rings would be unlinked, as they can be smoothly disengaged from the cluster.
However, since they are rigid, they are trapped between other rigid rings, remaining as an integral part of the clusters. We argue that these peculiar topological states can be distinguished by replacing, when treating rigid rings, isotopies -- deformations -- with isometries -- rotations and translations -- in the definition of link equivalence.
The results reported herein indicate that, by factoring in cut-and-saw operations, like reconnection, a mixture of polymers and threadable ring colloids, or sufficiently rigid semi-flexible rings (see~\cite{bonato2025prl}), can be transformed into a connected cluster of rings with tunable topological complexity.


\section{Conclusions}

In summary, in this work we have shown that the architecture, topology, and mechanical function of topological gels and bi-gels can be dynamically programmed \emph{post-assembly} without modifying their underlying microscopic constituents or chemical synthesis. By focussing mostly on a mixture of rigid colloidal rings and flexible linear polymers, which gives rise to a bi-gel, we show that the post-assembly mobility of entanglements and their anchoring at the boundary, confinement geometry, and dynamical reconnection function as three independent physical levers to reconfigure soft topological composite materials on demand.

First, we have established that the spatial mobility of topological entanglements acts as a post-assembly thermodynamic control parameter. By switching boundary anchors from quenched (fixed) to annealed (sliding) after network formation, the material transitions from a static, load-bearing elastic matrix into an adaptive network capable of internal stress redistribution and sustained, work-generating expansion (which grows diffusively over time). The mechanical response is therefore governed not only by the density of the entanglements but also by their dynamics.

Second, we have shown that changes in confinement geometry serve as a template to program gel morphology and topology, systematically transforming isotropic 3D bulk gels into 2D surface-spanning nets and quasi-1D interlocked ring networks. We have formalised these architectural transitions through graph-spectral analysis, demonstrating that adjacency matrix spectra provide a fingerprint of the different morphologies of topological networks.

Finally, introducing post-assembly strand reconnection converts the rigid-ring scaffold into an active topological rewrite engine. The rigid rings channel random reconnection events into elementary Hopf links ($|\text{Lk}| = 1$), suppressing the high-order knotting typical of pure polymer networks and creating {\it isometry-restricted trapped states} that are unlinked under continuous ambient isotopy, yet permanently interlocked under physical dynamics. 

Overall, these results suggest a general paradigm for programmable topological soft matter, in which topology is no longer viewed as a static structural property but as a dynamic design variable that can be manipulated through external protocols. Therefore we suggest that, rather than synthesising ever-more complex molecular building blocks, novel mechanical regimes can be programmed by directing where topological entanglements assemble, where they are permitted to slide, and how they are rewritten dynamically. We hope that these design principles may be applied in practice in the future to build topological matter via DNA-based catenated networks~\cite{tubiana2022}, mechanically interlocked molecular motifs~\cite{Polles2015}, 
slide-ring gels~\cite{ando2025}, or chromatin loop networks~\cite{dyson2021,bonato2024,chiang2025}. In particular, an interesting direction for future work is to investigate how activity~\cite{smrek2020active,ubertini2024universal,sakaue2026compounding,breoni2025,forte2026,dutta2026}, such as that generated by polymerases, and  DNA supercoiling~\cite{segers2025statistical,fosado2021,janissen2024,marko1995} would couple to our protocols, since energy-consuming enzymes and torsional stress could provide additional non-equilibrium levers on the mobility of the gel and rewriting of entanglements. Both could be realised by combining the two external protocols discussed here, anchoring and enzyme addition: for instance, supercoiling can be generated by adding gyrase to a network with quenched anchors, which prevent the torsional relaxation of the strands, whereas annealed anchors may allow it to dissipate.

\clearpage

\appendix
\section{Materials and methods}

We simulate a mixture of $N_r=90$ toroidal colloids and $N_p=30$ semi-flexible linear polymer chains.
These are modeled as bead-spring chains of $L_p = 108$ beads of size $\sigma$ held together by FENE potentials
\begin{equation}\label{UFENE}
U_{\rm FENE}(r) = \left\{
\begin{array}{lcl}
-0.5kR_0^2 \ln\left(1-(r / R_0)^2\right) & \ r\le R_0 \\ \infty & \
r> R_0 &
\end{array} \right. \, ,
\end{equation}
where $r$ is the distance between bonded beads, $k = 30\epsilon/\sigma^2$, and $R_{0}=1.5\sigma$.
The stiffness of the chains is modeled by introducing an energy penalty for large deviations from the equilibrium angles defined by consecutive triplets of neighboring beads. This penalty is described by the Kratky–Porod potential:
\begin{equation}\label{UKP}
U_{KP}(\theta) = \dfrac{k_BT l_p}{\sigma}[1-\cos(\theta)],
\end{equation}
where $l_p$ denotes the persistence length of the chains.
Each toroidal colloid is discretised as a ring of $\lfloor 2\pi R_r/\sigma\rfloor$ beads of size $\sigma$, where $R_r$ is the colloid radius, and is treated as a rigid body. By selecting a value for $R_r$, we can prepare the system either in the gel or sol phase (see~\cite{bonato2025prl}). In this work, we set $R_r=6.4\sigma$ for the gel phase, and $R_r=3.2\sigma$ for the sol phase.
All beads, whether belonging to rings or polymers, interact solely (excluding bonds and angular potentials) through steric repulsion, which is modeled using the truncated Lennard-Jones potential
\begin{equation}\label{UWCA}
U_{WCA}(r) = \left\{\
\begin{array}{lcl}
4\epsilon \left [ \left ( \frac{\sigma}{r}\right )^{12} - \left (\frac{\sigma}{r} \right )^6 \right ] + \epsilon & r \leq r_c\\
0 & r > r_c
\end{array} \right. \, ,
\end{equation}
where $r_c=2^{1/6}\sigma$, and $r$ is the distance between interacting beads.

As detailed in the results sections, throughout our simulations, we make use of either attractive or repulsive surfaces. We consider either spherical surfaces or planar walls. For most simulations, the system is assembled and equilibrated inside a sphere of radius $R_S=25\sigma$.
Beads interact with a repulsive surface via the potential,
\begin{equation}\label{CONF}
U_{Wall}(r) = \left\{\
\begin{array}{lcl}
\infty & R_S - r \leq 0 \\
\epsilon \left[\frac{2}{15}\left(\frac{\sigma}{R_S-r}\right)^9 - \left(\frac{\sigma}{R_S-r}\right)^3 \right] + \epsilon\frac{\sqrt{10}}{3}& 0 < R_S - r \leq r_{c,wall}\\
0 & R_S - r > r_{c,wall}
\end{array} \right. \, ,
\end{equation}
where $R_S$ is the radius of the sphere or the position of the wall (assuming it is perpendicular to one of the main axes), $r_{c,wall} = \sigma \left(\frac{2}{5}\right)^{1/6}$, and $r$ is now the distance from the surface.
To model attractive surfaces, we use the Morse potential:
\begin{equation}\label{CONF_ATT}
U_{Morse}(r) = 22\epsilon \left( e^{-2\alpha(r-r_0)}-2e^{-\alpha(r-r_0)} \right) - 22\epsilon \left( e^{-2\alpha(r_c-r_0)}-2e^{-\alpha(r_c-r_0)} \right), \qquad r < r_c
\end{equation}
where $r$ is the distance from the surface and $r_c$ a cutoff. We use $\alpha = 2.5\sigma^{-1}$, $r_0 = 0.2\sigma$ and $r_c = 2.0\sigma$.

The equations of motion are integrated in LAMMPS~\cite{LAMMPS} using the NVE ensemble coupled to a Langevin thermostat, with a timestep $dt=0.01\tau_{LJ}$ where $\tau_{LJ}= \sqrt{m\sigma^2/\epsilon}$.
Every set of simulations is averaged over 20 independent replicas.

To model bond cleavage, we use the LAMMPS fix bond/break command, which attempts to remove any bond with a certain probability $p_c$ at regular intervals $\Delta t_c$, provided the bond length is larger than a specific cutoff set to $0.5\sigma$. To extract the reaction rate, we fit the number of remaining bonds over time to $N_{bnd}(t)= N_{bnd}(0)e^{-k_ct}$. For $p_c=6\cdot10^{-4}$ and $\Delta t_c=1\tau_{LJ}$ our fit yields $k_c=(2.764\pm0.006)\cdot10^{-4}\tau^{-1}_{LJ}$.

For certain simulation sets, we also allow reconnections of the polymers, whereby the bonds of two nearby polymer segments are occasionally swapped. As described in Ref.~\cite{bonato2022}, this is implemented through a slightly modified version of the bond/swap fix in LAMMPS~\cite{Karayiannis2002PRL,Auhl2003}. To ensure detailed balance, reconnection attempts are accepted or rejected according to a Metropolis test.

\subsection{Anchoring protocols}

\begin{figure}[h!]
	\includegraphics[width=0.7\textwidth]{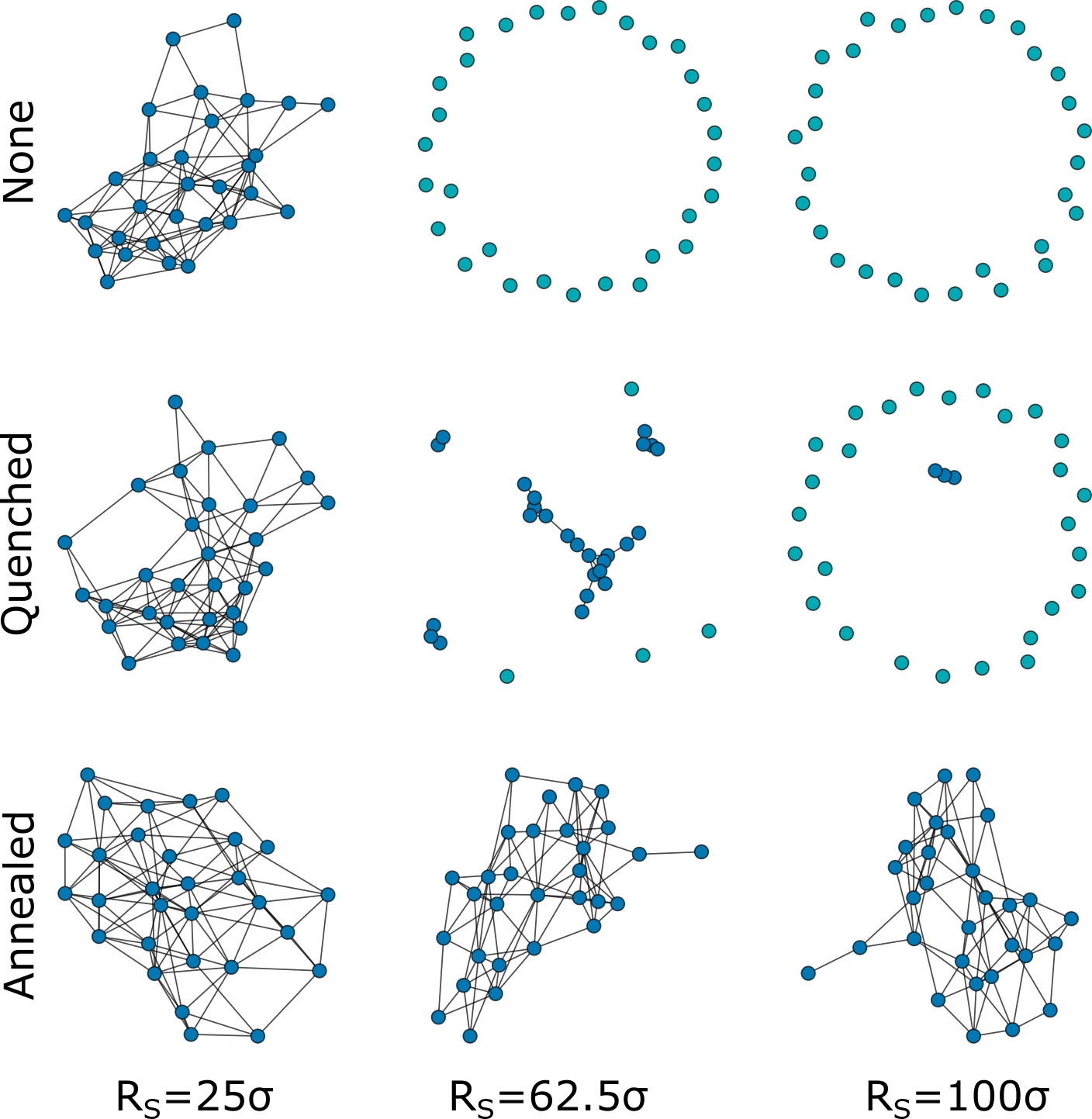}
    \caption{{\bf Network representation of the bi-gel.} 
    Polymer network representation with different anchoring styles (none, quenched, and annealed) for three values of the confinement radius $R_S=25\sigma$, $R_S=62.5\sigma$, and $R_S=100\sigma$.
    }
\label{fig_anch_net_SI}
\end{figure}

Before expansion we anchor the ends of the polymers to the confining surfaces. Two anchoring modes have been considered: (i) annealed anchoring, where the ends can slide on the surface, and (ii)  quenched anchoring, in which the polymer ends are kept fixed in position during the dynamics. Annealed anchoring is achieved by first superimposing on the confining sphere another sphere interacting attractively through a Morse (Eq.~\eqref{CONF_ATT}) potential with the ends of the polymers. Concurrently, we slightly reduce the radius of confinement of the rings, so that they are unable to unthread any threaded polymer. Since this potential depends only on the distance from the anchored boundary, once the ends bind to the surface, they are free to slide along it (annealed anchoring). To transform the anchoring to quenched, we let the ends bind and equilibrate, then we remove the attractive sphere and place some attractive traps -- beads interacting via Morse -- next to the ends. To incorporate these attractive traps into the expanding confinement sphere, we implemented a LAMMPS fix that moves them radially at a rate matched to the decompression rate.

\subsection{Network analysis of the topological complexity of the bi-gel }

The topological state of the system can be represented as a network of $N_p$ nodes, where each node corresponds to one linear polymer in the mixture. Two nodes are connected by a bond if there is at least one ring threaded by both corresponding polymers. The adjacency matrix of the network, $A_{ij}$,
\begin{equation}
A_{ij} = \begin{cases}
1\quad \text{$i$ and $j$ are connected} \\
0\quad \text{otherwise} \\
\end{cases}, 
\end{equation}
can be exploited to identify the connected components of the networks, associated with clusters of linked polymers (for more details, see Ref.~\cite{bonato2025prl}). Furthermore, as explained in the main text, the spectra of the adjacency matrix of the largest connected component of the system, obtained by eliminating rows and columns associated with polymers not belonging to the cluster, can be used to infer structural differences between gels assembled with different protocols. Fig.~\ref{fig_anch_net_SI} shows examples of the network representation of configurations sampled from simulations of spherical decompression without anchoring and with annealed and quenched anchoring.  In particular, we show how the network evolves as the confinement sphere expands (i.e., as the radius of the confinement sphere increases).



\bibliography{designing_top_mm}

\end{document}